\documentclass[a4paper,UKenglish,cleveref, autoref, thm-restate]{lipics-v2021}
\pdfoutput=1 %uncomment to ensure pdflatex processing (mandatatory e.g. to submit to arXiv)
\hideLIPIcs  %uncomment to remove references to LIPIcs series (logo, DOI, ...), e.g. when preparing a pre-final version to be uploaded to arXiv or another public repository

\usepackage{graphicx}
\usepackage{listings}
\usepackage{float} %needed for making listings float as figures
\newfloat{lstfloat}{htbp}{lop}
\floatname{lstfloat}{Listing}
\usepackage{textcomp}
\usepackage{xspace}
\usepackage{upgreek}
\usepackage{makecell}

\usepackage[utf8]{inputenc}

\usepackage{booktabs}
\usepackage{comment}
\usepackage{amsmath}
\usepackage{amssymb}
\usepackage{url}
\usepackage{hyperref}
\usepackage{cleveref}
\usepackage{ifthen}
\usepackage{soul}
\usepackage{multirow}
\usepackage{rotating}

\usepackage{makecell}
\usepackage{subcaption}
\usepackage{fancyvrb}
\usepackage[most]{tcolorbox}
\usepackage{arrayjobx}
\usepackage{svg}
\usepackage{algorithm}
\usepackage{algorithmic}
\usepackage[colorinlistoftodos]{todonotes}

\newboolean{showcomments}
\setboolean{showcomments}{true}
\ifthenelse{\boolean{showcomments}}
{
	\definecolor{myyellow}{RGB}{255, 228, 26}
	\definecolor{myblue}{RGB}{50, 50, 220} 
	\newcommand{\nb}[2]{
		{\sf
			\fcolorbox{myyellow}{yellow}{\scriptsize\textbf{#1}}%
			$\blacktriangleright$%
			{\color{myblue}\fontsize{7pt}{8pt}\selectfont\textbf{#2}}%
		}%
	}
}
{
	\newcommand{\nb}[2]{}
}

\definecolor{darkgreen}{rgb}{0,0.52,0}

\newcommand{\head}[1]{\noindent\textbf{#1.}}
\newcommand{\changed}[1]{\textcolor{black}{#1}}
\newcommand{\camera}[1]{\textcolor{black}{#1}}

\newif\ifincludeRQThree
\includeRQThreefalse
\newcommand{\withRQthree}[1]{\ifincludeRQThree #1\fi}

\author{Xingcheng Chen\footnote{Corresponding author}}{Technical University of Munich, fortiss, Germany}{xingcheng.chen@tum.de}{[0009-0002-0861-4093]}{}

\author{Mehmet Besenk}{Technical University of Munich, Germany}{mehmet.besenk@tum.de}{}{}

\author{Andrea Stocco}{Technical University of Munich, fortiss, Germany}{andrea.stocco@tum.de}{[0000-0001-8956-3894]}{}

\authorrunning{X. Chen et al.} %TODO mandatory. First: Use abbreviated first/middle names. Second (only in severe cases): Use first author plus 'et al.'



\nolinenumbers
\EventEditors{Robert Feldt and Maria Paasivaara}
\EventNoEds{20}
\EventLongTitle{20th International Symposium on Empirical Software Engineering and Measurement (ESEM 2026)}
\EventShortTitle{ESEM 2026}
\EventAcronym{ESEM}
\EventYear{2026}
\EventDate{October 4--9, 2026}
\EventLocation{Munich, Germany}
\EventLogo{}
\SeriesVolume{42}
\ArticleNo{23}
\begin{document}

\title{Explanation-Guided Metamorphic Testing of Specialized Language Models: An Empirical Study}

% \author{Anonymous Authors}{A}{}{}{}

\maketitle
% !TEX root =  paper.tex
\begin{abstract}
\head{Background} Task-specialized language models are increasingly integrated into software engineering workflows to support vertical-domain activities such as issue triaging, document classification, and automated analysis. Despite their adoption, there is limited empirical evidence on how to test their robustness and detect brittle behaviors under semantics-preserving input transformations.

\head{Aims} This paper investigates whether explainability-guided metamorphic testing can improve the effectiveness and validity of robustness testing for specialized language models compared to heuristic mutation strategies.

\head{Method} We conduct a large-scale empirical study of explanation-guided metamorphic testing across three datasets, four model architectures, and 20 testing configurations derived from combinations of attribution methods and mutation strategies. The evaluated configurations combine attribution-based token prioritization, LLM-driven mutation, and automated semantic verification to generate linguistically valid test variants. We assess failure discovery capability, semantic validity, and testing efficiency against heuristic baselines.

\head{Results} Explanation-guided metamorphic testing generates 2.30$\times$ more verified failure-inducing test cases than heuristic mutation strategies. 
\changed{Semantic verification substantially improves mutation validity and achieves high label-preservation precision among gate-accepted variants according to human annotation.}
The study further reveals systematic shortcut behaviors across models, including over-reliance on named entities and formatting cues. 
\withRQthree{In addition, adversarial refinement using verified variants improves robust accuracy by up to 7.4\% without substantially degrading performance on the original tasks.}

\head{Conclusions} The results provide evidence that explanation-guided metamorphic testing is an effective and practical approach for empirically evaluating \withRQthree{and improving} the robustness of task-specialized language models used in vertical AI applications.
\end{abstract}

\section{Introduction}\label{sec:introduction}

\noindent
Large language models (LLMs) are increasingly integrated into software engineering (SE) workflows, supporting tasks such as issue triaging, document classification, and automated analysis~\cite{liang2023holisticevaluationlanguagemodels}. Beyond large proprietary foundation models, 
there are many task-specialized language models adapted for a specific downstream task and
deployed as components of vertical AI applications. These models may be obtained by distillation, fine-tuning, or lightweight adaptation of general pretrained backbones, to achieve lower latency and deployment cost while maintaining strong task-specific performance.

\changed{Despite their growing adoption, specialized language models remain highly vulnerable to shortcut learning, spurious lexical correlations, and brittle decision boundaries inherited from fine-tuning datasets. Small semantics-preserving perturbations, such as modifying entity names, formatting cues, or sentiment-bearing expressions, can therefore trigger inconsistent predictions and unreliable downstream behavior. Existing robustness testing approaches rely on black-box perturbation strategies or manually designed transformations, which often generate unrealistic or semantically invalid test cases while providing limited insight into why failures occur~\cite{10.1145/3641289}. 
We therefore study robustness in settings representative of practical vertical AI deployments, where compact fine-tuned models are commonly preferred over large proprietary foundation models due to latency, privacy, and deployment constraints. Moreover, specialized language models are often deployed with full white-box access, enabling the use of attribution signals and internal explanations to identify decision-critical input regions~\cite{zhao2024explainability}. Despite this opportunity, there is still limited empirical evidence on whether explainability-guided testing can systematically improve the generation of valid failure-inducing test cases and expose shortcut-driven robustness weaknesses in specialized language models.}

Prior work has shown that language models frequently rely on brittle decision boundaries and shortcut behaviors, including superficial lexical artifacts, entity biases, and formatting cues~\cite{gardner-etal-2020-evaluating,jin2020textfooler,ribeiro-etal-2020-beyond}. Traditional evaluation pipelines mainly rely on static benchmark datasets that act as fixed test suites and primarily measure average generalization performance~\cite{wang2019gluemultitaskbenchmarkanalysis}. Such evaluations provide limited support for systematically generating corner cases capable of exposing hidden robustness weaknesses. Manual construction of challenging test cases is costly and difficult to scale~\cite{kiela-etal-2021-dynabench}, while existing automated mutation strategies often generate semantically invalid or unrealistic perturbations.

In this paper, we present an empirical study of explainability-guided metamorphic testing for task-specialized language models. The study investigates whether attribution-based guidance can improve the effectiveness and validity of robustness testing compared to heuristic mutation strategies. To this end, we evaluate combinations of explainable AI (XAI) attribution, LLM-driven mutation, and semantic verification mechanisms that generate targeted metamorphic variants while preserving semantic consistency. Attribution analysis identifies decision-critical tokens, a locally deployed Judge LLM produces controlled ablation and adversarial injection variants, and a semantic validation stage based on bidirectional natural language inference (BNLI) and log-perplexity shift (LPS) filters invalid mutations.

We conduct an empirical evaluation across multiple model architectures, including DistilBERT, BART-large, and Qwen-0.5B, and across diverse classification tasks, including SST-2, AG News, and GitHub Issue classification~\cite{socher2013sst,zhang2015agnews}. Our results show that explanation-guided metamorphic testing generates substantially more verified failure-inducing test cases than heuristic mutation strategies while reducing semantically invalid perturbations. Furthermore, our analysis reveals systematic shortcut behaviors, including entity over-reliance in discriminative models and sensitivity to formatting artifacts in generative models. \withRQthree{Finally, we show that verified metamorphic variants can support adversarial refinement, improving robust accuracy by up to 7.4\% without substantially degrading performance on the original tasks.}

\section{Background}\label{sec:background}

\subsection{Task-Specialized Language Models}

LLMs are commonly built on Transformer architectures~\cite{vaswani2017attention} and include encoder-only models (e.g., BERT~\cite{devlin2019bert}), encoder--decoder models (e.g., BART and T5~\cite{bart,raffel2020t5}), and decoder-only models (e.g., GPT-style models and Qwen~\cite{qwen}). In practical software engineering workflows, however, deploying large proprietary models is often infeasible due to latency, cost, and privacy constraints. As a result, many SE applications rely on \emph{task-specialized language models} obtained through distillation or supervised fine-tuning of smaller architectures~\cite{DistilBERTAD,qwen}.

While these models provide improved efficiency and deployability, prior work has shown that language models frequently rely on brittle decision boundaries and shortcut behaviors, including lexical artifacts, entity biases, and formatting cues~\cite{ribeiro-etal-2020-beyond,gardner-etal-2020-evaluating,jin2020textfooler}. Such weaknesses can compromise robustness in production SE pipelines, motivating the need for systematic robustness testing approaches.

\subsection{Metamorphic Testing}

Metamorphic Testing (MT) addresses the \emph{oracle problem} by validating the consistency of a system's behavior across related inputs instead of requiring a ground-truth output for every test case~\cite{chen1998metamorphic,segura2016survey}. MT defines \emph{metamorphic relations} (MRs), which specify how outputs should remain invariant, or change predictably, under controlled input transformations.

For language models, typical MRs involve semantics-preserving transformations such as paraphrasing, synonym substitution, or contextual edits that should not alter the expected label. By generating related inputs and comparing predictions, MT can expose robustness violations and shortcut behaviors.

Recent work has explored metamorphic testing for language models through manually defined or heuristic transformations~\cite{11185922}. In this paper, we empirically investigate whether explainability-guided mutation and semantic verification improve the effectiveness and validity of metamorphic testing for task-specialized language models.

\subsection{Explainable AI for Language Models}

Explainable AI (XAI) techniques estimate the contribution of individual input features to a model's prediction~\cite{zhao2024explainability}. In language models, this typically corresponds to assigning importance scores to input tokens. Attribution methods generally fall into three categories: (1)~gradient-based methods, such as Integrated Gradients~\cite{sundararajan2017integrated,mudrakarta2018did}; (2)~perturbation-based methods, such as Occlusion and LIME~\cite{zeiler2014occlusion,li2016erasure,ribeiro2016lime}; and (3)~architecture-aware methods based on Transformer attention patterns~\cite{abnar2020attention}.

Although XAI techniques are primarily used for model interpretation, attribution scores can also support robustness testing by identifying decision-critical tokens. Prior studies suggest that highly influential tokens often correspond to shortcut features or brittle lexical cues~\cite{ribeiro-etal-2020-beyond}. In this work, we empirically study whether attribution-guided mutations are more effective than unguided transformations at generating valid failure-inducing test cases.
\section{Empirical Study}\label{sec:approach}

This study evaluates combinations of attribution methods, metamorphic mutation strategies, and semantic verification mechanisms for generating failure-inducing test cases.

The empirical evaluation investigates three methodological aspects of metamorphic robustness testing:
(1)~attribution-based mutation prioritization,
(2)~metamorphic mutation strategies, and
(3)~semantic verification, including human validation of the automatic gate.

\subsection{Attribution-Based Mutation Prioritization}

The study first investigates whether attribution scores can effectively guide mutation selection toward decision-critical input regions. Rather than relying on a single explanation technique, we evaluate multiple attribution strategies as interchangeable experimental factors.

Formally, given an input sequence $x = [t_1, \dots, t_n]$, the attribution process computes a saliency vector
$\mathbf{a} = [a(t_1), \dots, a(t_n)]$, where $a(t_i)$ denotes the influence of token $t_i$ on the model prediction.

\head{Gradient-based Attribution}
When gradient access is available, we evaluate Integrated Gradients (IG)~\cite{sundararajan2017integrated}, defined as:

\begin{equation}
a_{ig}(t_i) = (e_i - e_i^0)
\int_{0}^{1}
\frac{\partial f(x^0 + \alpha(x-x^0))}{\partial e_i} d\alpha
\end{equation}

where $e_i$ and $e_i^0$ denote the embedding vectors of token $t_i$ and its baseline representation.

\head{Perturbation-based Attribution}
We also evaluate Occlusion~\cite{zeiler2014occlusion}, which estimates token importance by measuring prediction changes when a token is masked:

\begin{equation}
a_{occ}(t_i) = f(x) - f(x \setminus \{t_i\})
\end{equation}

where $x \setminus \{t_i\}$ replaces token $t_i$ with a baseline token such as \texttt{[PAD]}.

\head{Architecture-aware Attribution}
For Transformer-based models, we additionally evaluate attention-based attribution methods~\cite{abnar2020attention,chefer2021transformer}, which estimate token importance using attention aggregation and gradient-weighted attention scores.

Since many language models operate on subword tokens, token-level attribution scores are aggregated at the word level, i.e., $S(w_j) = \sum_{t \in w_j} a(t)$.
The resulting ranked word set
$\mathcal{T}(x) = \{(w_j, S(w_j))\}$
defines the mutation priority used during metamorphic mutation.

\begin{table*}[t]
\centering
\caption{Illustrative examples of metamorphic mutations. Tokens with high attribution scores are underlined. MR-Ablation weakens influential lexical cues, while MR-Injection introduces adversarial context. Semantic equivalence is validated through the NLI score ($\mathcal{M}_{nli}$) and linguistic naturalness through the perplexity shift metric ($\mathcal{M}_{lps}$).}
\label{tab:examples}
\resizebox{\textwidth}{!}{
\begin{tabular}{lp{5.2cm}cp{5.2cm}cc}
\toprule
\textbf{MR Type} & \textbf{Original Input ($x$)} & \textbf{Saliency} & \textbf{Mutated Variant ($x'$)} & \textbf{$\mathcal{M}_{nli} \uparrow$} & \textbf{$\mathcal{M}_{lps} \downarrow$} \\
\midrule
\textit{Ablation} & The \underline{acting} was \underline{superb}, but the plot was weak. & superb (0.82) & The acting was good, but the plot was weak. & 0.74 & 0.05 \\
\textit{Ablation} & A \underline{boring} and \underline{predictable} movie. & boring (0.75) & A plain and predictable movie. & 0.81 & 0.12 \\
\textit{Injection} & The movie is a masterpiece of cinema. & disaster (injected) & The movie is a masterpiece, despite being a disaster to some. & 0.78 & 0.18 \\
\textit{Injection} & It provides a fresh perspective. & terrible (injected) & It provides a fresh, albeit terrible, perspective. & 0.65 & 0.25 \\
\bottomrule
\end{tabular}
}
\end{table*}

\subsection{Metamorphic Mutation Strategies}

Given an input text $x$ with label $y$ and a specialized language model as the system under test (SUT) $f(\cdot)$, the evaluation seeks to generate metamorphic variants $\mathcal{X}'=\{x'_i\}$ such that: (i)~$x'_i$ remains semantically equivalent to $x$, and (ii)~the model prediction changes ($f(x'_i)\neq f(x)$). Such inconsistencies indicate brittle decision boundaries, shortcut behaviors, or spurious correlations.
The study next evaluates different strategies for generating metamorphic variants while preserving semantic meaning. Compared to rule-based perturbations, LLM-driven mutations enable context-aware rewrites that remain linguistically natural.

\autoref{tab:examples} presents representative mutation examples used throughout the study. We evaluate two primary metamorphic relations (MRs).

\head{MR-Ablation}
This relation evaluates over-reliance on influential lexical features. Given an input $x$, the mutation targets the word $w \in \mathcal{T}(x)$ with the highest attribution score. The selected token is either removed or replaced with a semantically weaker alternative while preserving grammaticality.

\head{MR-Injection}
This relation evaluates robustness against distracting lexical cues. Adversarial trigger words extracted from misclassified examples are inserted into the input while preserving semantic coherence and fluency.

Mutations are generated using a locally deployed Judge LLM and processed in batches to improve efficiency. Each generated variant is accompanied by a structured self-verification output indicating whether semantic and grammatical constraints are satisfied.

\subsection{Semantic Verification}

A major challenge in generating metamorphic variants is the risk of producing invalid test cases, i.e., mutations that alter the underlying task label and lead to false positives. To mitigate this risk and establish a reliable, yet approximated automated oracle, we implement a validation stage that filters variants based on semantic consistency and linguistic naturalness.

\head{Bidirectional Semantic Entailment}
To verify that the mutation preserves the semantic meaning of the original input $x$, we employ a cross-encoder Natural Language Inference (NLI) model~\cite{williams2018mnli,liu2019roberta}. 
However, uni-directional entailment ($x \rightarrow x'$) is insufficient for testing: while it ensures the mutation $x'$ does not introduce contradictory information, it does not penalize the removal of important context. Conversely, the reverse relation ($x' \rightarrow x$) verifies that the mutated input still entails the constraints expressed in the original sentence.
We therefore enforce a \emph{bidirectional} mutual entailment score:

\[
\mathcal{M}_{nli}(x, x') =
\min \left(
P(E \mid x, x'),
P(E \mid x', x)
\right)
\]

where $P(E)$ denotes the probability of entailment predicted by the NLI model. 
A mutation is accepted only if $\mathcal{M}_{nli}(x, x') > \tau_{nli}$. 
This bidirectional constraint reduces the likelihood that a mutation introduces semantic drift or removes essential contextual information.

\head{Linguistic Naturalness Metric}
While NLI captures semantic consistency, NLI models are often tolerant to grammatical errors or unnatural phrasing. As a result, they may assign high entailment scores to syntactically degraded inputs. If the SUT fails on such inputs, the failure may stem from out-of-distribution syntax rather than a genuine logical vulnerability.

To filter these artifacts, we evaluate linguistic fluency using perplexity~\cite{jozefowicz2016exploring}, defined as the exponentiation of cross-entropy loss $\mathcal{L}(\cdot)$. 
Rather than using absolute perplexity values, we measure the \emph{log-perplexity shift} between the original and mutated inputs:

\[
\mathcal{M}_{lps}(x, x') =
\log \frac{\text{PPL}(x')}{\text{PPL}(x)} 
%= \log \text{PPL}(x') - \log \text{PPL}(x) 
= \mathcal{L}(x') - \mathcal{L}(x)
\]

This metric isolates the fluency degradation introduced by the mutation itself. By subtracting the baseline loss of the original input, the measure reduces the influence of the intrinsic difficulty of the original sentence and focuses on the perturbation introduced by the metamorphic edit.
A mutation $x'$ is rejected if $\mathcal{M}_{lps}(x, x') > \tau_{lps}$, reducing the likelihood that accepted variants are linguistically unnatural or far from the original distribution.

\subsection{Research Questions}

\noindent
\textbf{RQ\textsubscript{1} (configuration sensitivity ).}
\textit{How do different attribution, mutation, and semantic verification configurations affect the validity and failure discovery capability of metamorphic robustness testing?}

The study evaluates multiple configurable components involved in the generation and verification of metamorphic variants, including attribution mechanisms, mutation strategies, and semantic filtering criteria. 

\noindent
\textbf{RQ\textsubscript{2} (effectiveness and efficiency).}
\textit{How effective and efficient are attribution-guided metamorphic testing configurations at discovering prediction inconsistencies across different models and datasets?}

To assess the practical applicability of our study, we evaluate the capability of attribution-guided metamorphic testing to expose robustness violations while maintaining reasonable computational overhead across diverse architectures and classification tasks.

\noindent
\textbf{RQ\textsubscript{3} (human validation).}
\textit{To what extent does the automated semantic verification gate agree with
human judgments of label preservation and fluency?}

This RQ assesses whether variants accepted by the semantic gate are judged
valid by human annotators. It measures agreement between automatic gate
decisions and human annotations for label preservation and fluency.

\subsection{Metrics}

For RQ\textsubscript{1} and RQ\textsubscript{2}, we assess each testing configuration using two metrics: (i) the \emph{Attack Success Rate (ASR)}, defined as the proportion of generated metamorphic variants that change the SUT prediction, and (ii) the number of \emph{verified failures}, i.e., prediction-changing variants satisfying the semantic verification criteria. We also report the average time required to generate the first verified failure as a measure of efficiency.

For RQ\textsubscript{3}, we evaluate semantic-gate quality against human judgments. We
report confusion statistics between automatic gate decisions and majority
human labels, including accuracy and pass precision. Pass precision denotes
the proportion of gate-accepted variants that are judged valid by humans.

\withRQthree{For RQ\textsubscript{3}, we evaluate the impact of adversarial refinement using two criteria.
First, we measure post-refinement accuracy on the original test sets and on external adversarial benchmarks to assess potential degradation on in-distribution data.
Second, we report the reduction in ASR after refinement, which quantifies the robustness improvement against semantically preserving metamorphic variants.}

\subsection{Objects of Study} 

\subsubsection{Datasets} 

To assess generalizability across diverse linguistic complexities and application domains, we adopted three text classification benchmarks, widely used in %previous research
\camera{SE testing work that evaluates LLM testing techniques on both domain-specific and standard classification tasks
~\cite{wang2019gluemultitaskbenchmarkanalysis, 11185922, METAL}.}

\head{SST-2}
The Stanford Sentiment Treebank (SST-2)~\cite{socher2013sst} is a widely adopted GLUE benchmark consisting of 67k human-annotated movie reviews for binary sentiment analysis. We select SST-2 because it is notoriously rich in complex syntactic phenomena, such as contrastive conjunctions and nuanced negation. It serves as an ideal testbed to evaluate whether our metamorphic mutations effectively expose vulnerabilities in the model's compositional understanding.

\head{AG News}
For general topic classification, we employ the AG News dataset~\cite{zhang2015agnews}, containing 120k training samples categorized into four distinct news domains (World, Sports, Business, Sci/Tech). This dataset evaluates the model's reliance on broad vocabulary and named entities, acting as a standard control to verify that the Judge LLM can perform adversarial injections without drifting across strict topical boundaries.

\head{GitHub Issue Classification}
We utilize the large-scale GitHub Issue dataset originally curated for the NLBSE tool competition~\cite{nlbse2022} and further refined for LLM fine-tuning tasks. This dataset consists of over 800k real-world issue titles and descriptions labeled into three categories: Bug, Enhancement, and Question. The complexity of this dataset stems from two primary factors: Multilingualism, as the issues are sourced from global open-source communities, containing technical discussions in multiple languages (e.g., English, Chinese, German, and Japanese); and Structural Noise, as the text heavily interweaves natural language with code snippets, URLs, and stack traces. Classifying these issues requires the SUT to perform deep logical reasoning and resist linguistic interference rather than relying on superficial keyword matching.

\subsubsection{Models}

\begin{table}[t]
\centering
\caption{Details of the SUTs and their original performance on the target datasets. Accuracies (\%) are reported on the standard test/validation splits after supervised fine-tuning.}
\label{tab:sut}
\resizebox{0.8\columnwidth}{!}{
\begin{tabular}{clccccccc}
\toprule
\multirow{2}{*}{\textbf{Model}} & \multirow{2}{*}{\textbf{Architecture}} &  \multirow{2}{*}{\textbf{Params}}  & \multicolumn{3}{c}{\textbf{Original Accuracy (\%) $\uparrow$}} \\ \cmidrule(lr){4-6}
 &  &  & \textbf{SST-2} & \textbf{AG News} & \textbf{GitHub} \\ \midrule
\makecell{DistilBERT-\\base-uncased \cite{DistilBERTAD}} & Encoder-only & 66M &  91.17 & 92.22 & 85.63 \\
\makecell{Qwen/Qwen2.5\\(Discriminative) \cite{qwen}} & Decoder-only & 0.5B & 92.20 & 93.93 & 86.28 \\
\makecell{Qwen/Qwen2.5-\\(Generative) \cite{qwen}} & Decoder-only & 0.5B & 93.69 & 91.95 & 83.39 \\
\makecell{facebook/bart-large \cite{bart}} & Encoder-decoder & 406M & 95.76 & 95.09 & 86.76 \\
\bottomrule
\end{tabular}
}
\end{table}

To evaluate generalizability across model architectures and inference paradigms, we selected four representative Transformer-based models spanning Encoder-only, Encoder--Decoder, and Decoder-only architectures, as well as both discriminative and generative classification settings.
Specifically, we evaluate \texttt{Qwen2.5-0.5B} under both discriminative and generative formulations to analyze how the inference paradigm influences susceptibility to metamorphic perturbations.
\changed{Although the evaluated architectures originate from general-purpose pretrained language models, the systems under test are fine-tuned for downstream classification tasks and deployed as task-adapted inference pipelines.}
The detailed model specifications and supervised fine-tuning accuracies are reported in \autoref{tab:sut}. 
All SUTs achieve strong baseline performance after fine-tuning, with accuracies ranging from 83.39\% to 95.76\% across the evaluated datasets. Among the evaluated models, \texttt{facebook/bart-large} achieves the highest overall accuracy.

\subsubsection{Configurations and Baselines}

To systematically analyze the impact of attribution and mutation choices, we define a configuration space of $4 \times 5 = 20$ setups obtained by combining four attribution mechanisms with five mutation strategies.

\head{Attribution mechanisms (4 variants)}
We compare three attribution-based token prioritization methods---\emph{Integrated Gradients} (IG)~\cite{sundararajan2017integrated}, \emph{Occlusion}~\cite{zeiler2014occlusion}, and attention-based attribution~\cite{abnar2020attention,chefer2021transformer}---against a random-selection baseline without saliency guidance.

\head{Mutation strategies (5 variants)}
We compare two LLM-based semantic mutation operators---\emph{LLM-Ablate} and \emph{LLM-Inject}---against three rule-based mutation strategies: \emph{In-situ Ablation} (direct token deletion), \emph{Prefix Injection} (trigger-token prepending), and \emph{Random Injection}.
In particular, the unguided setup without attribution prioritization or LLM-based rewriting resembles the invariance-style perturbation testing of CheckList~\cite{ribeiro-etal-2020-beyond}, illustrating the limitations of blind mutations and their tendency to introduce grammatical artifacts instead of semantically meaningful perturbations.
The combination of attribution-guided token selection with rule-based mutations mirrors the core intuition of TextBugger~\cite{li2019textbugger}, where salient input regions are preferentially perturbed.
Similarly, prefix-based trigger insertion follows the attack-generation principle of XAI-Attack~\cite{bayer-etal-2024-xai} and Universal Adversarial Triggers~\cite{wallace2019triggers}, where adversarial keywords are inserted at the beginning of the sequence to exploit positional sensitivity and shortcut behaviors.
Finally, combining attribution-guided token selection with direct in-situ deletion is inspired by  the Representation Erasure paradigm~\cite{li2016erasure}. This configuration isolates the effect of semantic rewriting and enables analysis of how semantic smoothing influences mutation validity and failure generation.

\head{Implementation}
To instantiate the Judge LLM ($\mathcal{J}$) for metamorphic generation, we deploy the \textit{gpt-oss-20b} model locally using the Ollama framework. 
Using a locally hosted open-weights model ensures experimental reproducibility, avoids data privacy concerns and usage costs associated with cloud-based APIs, and prevents issues related to API version drift.
The Judge LLM operates with a temperature of $T = 1.0$, enabling broader exploration of the mutation space. Potentially invalid variants are subsequently filtered by the two-stage semantic verification procedure. 
We compute the NLI scores using the \textit{cross-encoder/nli-deberta-v3-small} model. Unlike standard bi-encoders that process sentences independently, cross-encoders perform full attention over the concatenated input pair, yielding higher accuracy and fine-grained sensitivity for textual entailment tasks~\cite{he2021debertav3, morris2020textattack}. 
To calculate the perplexity, we utilize the standard \textit{GPT-2} model~\cite{radford2019language}. As a foundational causal language model, GPT-2 serves as a highly efficient, lightweight, and universally accepted baseline for evaluating textual naturalness and perplexity in adversarial generation frameworks~\cite{garg-ramakrishnan-2020-bae, qi2021openattack}. 

All experiments, including supervised fine-tuning (SFT), XAI attribution computation, and metamorphic test generation, were conducted for three datasets and four model architectures on a single NVIDIA L40S GPU. The complete experimental pipeline required approximately 180 GPU hours. 

\subsection{Setup}

\subsubsection{Data Partitioning} 
To prevent data leakage and ensure a rigorous evaluation, we enforce a strict data partition. Specifically, we hold out a fixed subset of 2,000 instances from the original training data exclusively to compute XAI attributions and construct the adversarial candidate pools $\mathcal{T}(x)$. 
\camera{Before mutation generation, we discard the instances that are not correctly classified by the corresponding SUT from $\mathcal{T}(x)$.}
This holdout subset is strictly excluded from the SFT phase of the SUT. The remaining majority of the training data is utilized for standard SFT, while the original development and test splits are kept entirely untouched to evaluate model convergence and measure the final robust accuracy.

\subsubsection{Supervised Fine-Tuning} 
To comprehensively evaluate the robustness of our target models, we adapt the fine-tuning pipeline to accommodate the fundamental architectural differences between discriminative and generative paradigms.

For discriminative settings, models are fine-tuned for standard sequence classification and appended with a linear classification head to output class probabilities.
For generative settings, the classification task is reformulated as conditional text generation via instruction-based prompting. We wrap the inputs in a standardized template (e.g., \textit{``Instruction: Classify the following GitHub issue... Category:''}). To prevent the model from optimizing for input reconstruction, we apply target masking on prompts and inputs. This ensures the cross-entropy loss is computed exclusively on the generated target tokens (e.g., \textit{``bug''}) and the sequence terminator.

To maintain experimental consistency and resource efficiency, all models undergo standard SFT. Following established best practices for ensuring fine-tuning stability~\cite{devlin2019bert, mosbach2020on}, we train the models using a learning rate of $2 \times 10^{-5}$ with a warm-up ratio of 10\%. To prevent overfitting, the training process is capped at 9 epochs (7 for generative architectures) and is strictly governed by an early stopping mechanism with a patience of 2 epochs.

\subsection{Evaluation Procedure}

For RQ\textsubscript{1}, we evaluate the effectiveness of different attribution methods for identifying failure-inducing mutation targets under the direct MR-Ablation setting, where attribution scores alone determine which token is perturbed. For each $(\text{Model}, \text{Explainer})$ pair, we measure the ASR under three ablation strategies: removing the token with the highest positive attribution score, removing the token with the strongest negative attribution score, and removing a randomly selected token as a baseline. Effective attribution methods are expected to produce substantially higher ASRs when perturbing positively attributed tokens compared to negatively attributed or randomly selected tokens, indicating that the explainer successfully identifies prediction-relevant regions.

To calibrate the semantic verification stage, we determine the thresholds of the entailment and perplexity filters using the Kneedle algorithm~\cite{kneed}, which identifies inflection points in the metric distributions. For the entailment metric $\mathcal{M}_{nli}$, the threshold $\tau_{nli}$ is selected at the point preceding the sharp decline in retained failure-inducing variants, balancing semantic strictness and sample retention. For the linguistic naturalness metric $\mathcal{M}_{lps}$, the threshold $\tau_{lps}$ is selected at the elbow of the cumulative distribution to remove highly unnatural perturbations while retaining the majority of linguistically plausible variants.

For RQ\textsubscript{2}, we instantiate the evaluation using the most effective attribution method identified in RQ\textsubscript{1} and compare attribution-guided mutation selection against random selection to isolate the contribution of explainability-guided targeting. We evaluate both the Attack Success Rate and the number of generated failure-inducing variants under two conditions: \emph{pre-verification}, which includes all generated mutations, and \emph{post-verification}, which includes only variants satisfying both semantic thresholds $\tau_{nli}$ and $\tau_{lps}$. 
\changed{We further perform statistical analyses for the ASR difference between the studied
configuration and its baseline. We report the mean paired difference, a non-parametric 95\% bootstrap confidence interval over paired blocks~\cite{tibshirani1993bootstrap}, a one-sided Wilcoxon signed-rank test to estimate the stability of observed effects~\cite{Wilcoxon1945}, and Cohen's $d$ as effect size~\cite{cohen1988statistical}. 
}

\withRQthree{For RQ\textsubscript{3}, we investigate whether verified metamorphic variants can improve robustness through adversarial refinement. Successful failure-inducing variants passing the semantic verification stage are incorporated into an adversarial fine-tuning procedure designed to encourage prediction consistency between original inputs and their verified semantic variants. To reduce the risk of catastrophic forgetting, the verified adversarial variants are combined with the original training data during refinement. We employ Low-Rank Adaptation (LoRA) to efficiently update model parameters with limited computational overhead. The refined models are then evaluated on both the original development sets and dedicated adversarial evaluation sets to measure robustness improvements while monitoring potential degradation in in-distribution performance.}

For RQ\textsubscript{3}, we conduct a human evaluation study to assess whether the semantic verification stage aligns with human judgments of mutation validity. We sample mutations from all evaluated tasks, drawn throughout different attribution methods and mutation strategies. 
After removing empty outputs and non-English examples, the final evaluation set contains 383 mutation pairs. Annotators recruited through Amazon Mechanical Turk~\cite{sorokin2008utility} are shown the task description, the original task label, the original input, and the mutated variant. They then evaluate whether the mutated input preserves the original label and whether it remains fluent and natural. Each annotation batch contains one attention-check question, and submissions failing the attention check are discarded. In total, we collect 92 submissions, of which 84 pass the quality filter, yielding 818 valid item-level annotations. 
A mutation is considered human-valid only if it is judged both label-preserving and fluent. 
\camera{The annotations show moderate inter-annotator agreement with Fleiss' $\kappa=0.50$ and  88.9\% majority agreement, which is acceptable given the subjective nature of semantic-preservation and fluency judgments.}
\section{Results}

\subsection{Configuration sensitivity (RQ\textsubscript{1})}
\begin{figure}[t]
  \centering
  \includegraphics[width=0.7\linewidth]{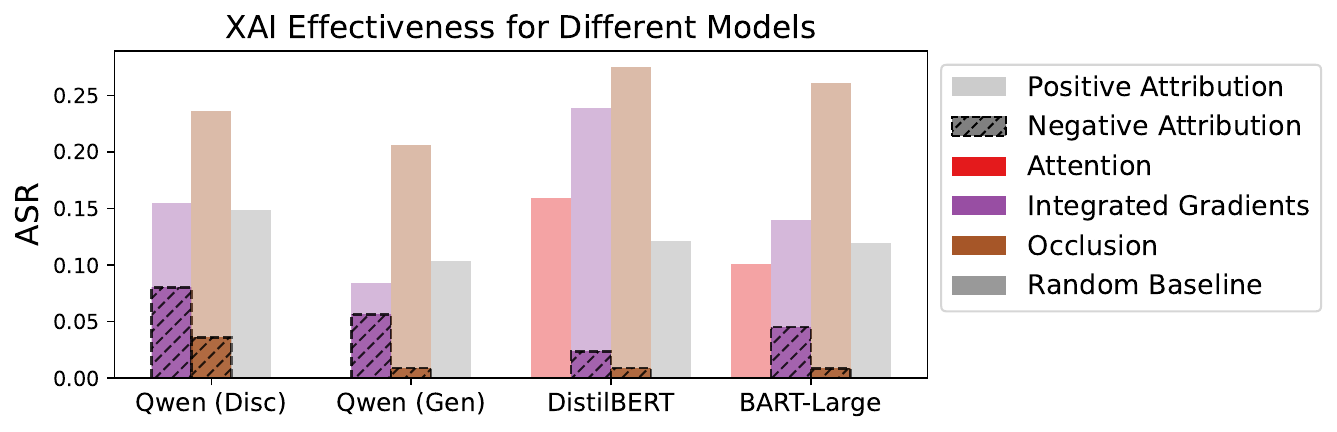}
  \caption{Effectiveness of different XAI explainers in guiding Metamorphic Ablation.}
  \label{fig:rq1_xai}
\end{figure}
\begin{figure*}[t]
  \centering
  \begin{subfigure}[t]{0.495\textwidth}
    \centering
    \includegraphics[width=\linewidth]{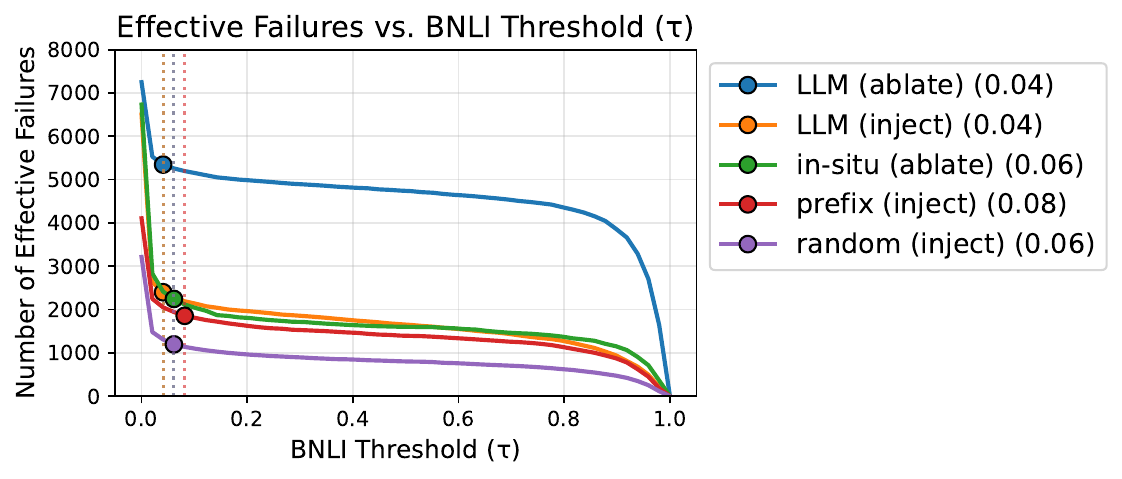}
    \caption{Failures across varying BNLI thresholds $\uparrow$}
    \label{fig:bnli-sweet-spot}
  \end{subfigure}
  \hfill
  \begin{subfigure}[t]{0.495\textwidth}
    \centering
    \includegraphics[width=\linewidth]{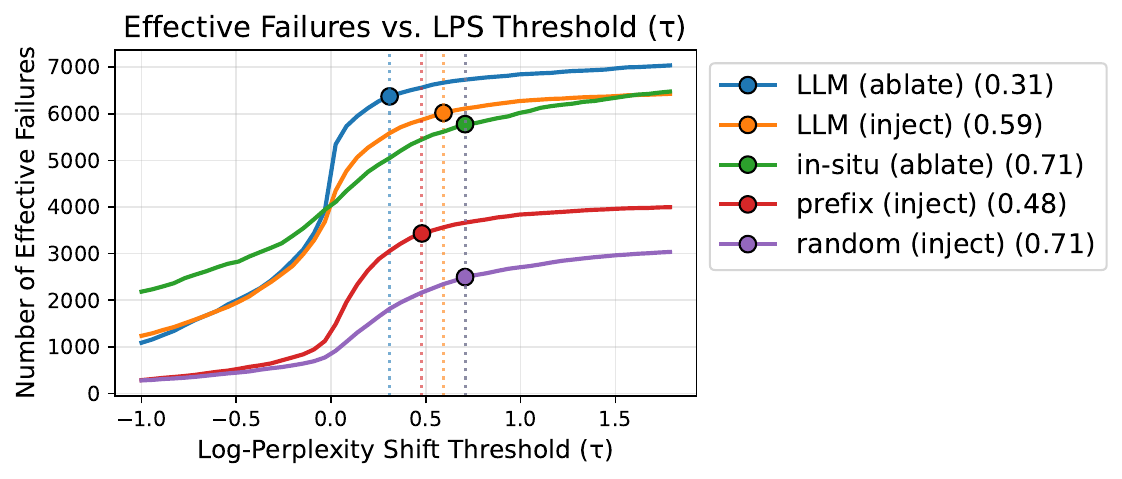}
    \caption{Failures across varying LPS thresholds $\downarrow$}
    \label{fig:lps-sweet-spot}
  \end{subfigure}
  \caption{Sweet-spot analysis for selecting automated verification thresholds (Kneedle algorithm). Circular markers ($\circ$) denote the knee-point thresholds ($\tau$) selected for each mutation strategy.}
  \label{fig:sweet_spot}
\end{figure*}

% \Andrea{Fig. 2 is cited before Fig. 1, fix}

\autoref{fig:rq1_xai} reports the Attack Success Rate obtained by different attribution methods under the direct MR-Ablation setting. Across all evaluated architectures, Occlusion consistently achieves the highest post-verification ASR values. In contrast, perturbing tokens with the strongest negative attribution scores (hatched bars) produces substantially lower ASRs. This clear separation provides an important sanity check, indicating that the attribution methods identify prediction-relevant regions rather than merely degrading sentence fluency through arbitrary token removal.

Among the evaluated explainers, Occlusion emerges as the most stable and architecture-agnostic strategy for identifying vulnerability-prone tokens. Its effectiveness remains consistent across encoder-only, encoder--decoder, and decoder-only architectures. Based on these results, we select Occlusion as the default attribution method for the remaining experiments.

\autoref{fig:sweet_spot} illustrates the number of retained failure-inducing variants under varying semantic verification thresholds. Across both semantic metrics, LLM-based semantic mutation strategies consistently produce substantially more semantically valid failures than rule-based baselines, indicating a higher ability to generate boundary-compliant perturbations while preserving linguistic plausibility.

To determine robust semantic verification thresholds, we apply the Kneedle algorithm to identify inflection points balancing semantic validity and sample retention. As shown in \autoref{fig:sweet_spot}(a), the number of retained failures decreases sharply as the bidirectional entailment threshold increases from $0.0$ to $0.1$, indicating that a large portion of raw perturbations exhibit semantic drift. Based on the detected inflection point, we select a unified threshold of $\tau_{nli}=0.08$, which filters semantically inconsistent variants while preserving the majority of semantically valid failures.

\autoref{fig:sweet_spot}(b) shows the cumulative distribution of log-perplexity shift values. The curve follows an S-shaped pattern, with inflection points located approximately in the range $\tau_{lps}\in[0.0,0.7]$. We select a unified threshold of $\tau_{lps}=0.3$, which removes highly unnatural perturbations while retaining most linguistically plausible variants.

\begin{tcolorbox}[boxrule=0pt,frame hidden,sharp corners,enhanced,borderline north={1pt}{0pt}{black},borderline south={1pt}{0pt}{black},boxsep=2pt,left=2pt,right=2pt,top=2.5pt,bottom=2pt]
\textbf{Configuration sensitivity (RQ\textsubscript{1})}: \textit{
Occlusion consistently provides the most effective attribution guidance.
Kneedle-based calibration identifies stable thresholds ($\tau_{nli}=0.08$ and $\tau_{lps}=0.3$) that balance semantic preservation and retained failure coverage.
}
\end{tcolorbox}

\begin{table*}[t]
\centering
\caption{Effectiveness of the evaluated mutation configurations across three datasets. We report the ASR results \textbf{Before/After} Semantic Verification Gate filtering. Best results after verification are marked in \textbf{bold}.}
\label{tab:combined_results}
\resizebox{\textwidth}{!}{
\begin{tabular}{ll @{\extracolsep{2pt}} cc cc cc cc}
\toprule 
\multirow{2}{*}{\textbf{MR}} & \multirow{2}{*}{\textbf{Strategy}} & \multicolumn{2}{c}{\textbf{Qwen-0.5B (Disc)}} & \multicolumn{2}{c}{\textbf{Qwen-0.5B (Gen)}}  & \multicolumn{2}{c}{\textbf{DistilBERT}} &  \multicolumn{2}{c}{\textbf{BART-Large}} \\
\cmidrule(r){3-4} \cmidrule(r){5-6} \cmidrule(r){7-8} \cmidrule(r){9-10}
& & Occlusion & Random  & Occlusion & Random   & Occlusion & Random & Occlusion & Random \\

\midrule

\multicolumn{10}{c}{\textbf{SST-2}} \\

\midrule

\multirow{2}{*}{\texttt{ablate}} & \texttt{lm} & 5.52/2.83 & 6.43/\textbf{3.35} & 4.30/\textbf{1.81} & 2.87/1.36 & 7.10/\textbf{2.54} & 4.72/2.28 & 5.67/\textbf{2.28} & 4.49/1.85 \\
 & \texttt{in-situ} & 15.40/2.79 & 14.83/2.55 & 12.73/0.86 & 10.39/0.94 & 17.57/1.71 & 12.10/1.17 & 18.31/1.51 & 11.99/1.37 \\
\cmidrule{1-10}
\multirow{3}{*}{\texttt{inject}} & \texttt{lm} & 9.57/0.85 & 8.40/0.53 & 6.68/0.52 & 4.44/0.31 & 8.70/0.48 & 6.31/0.21 & 6.76/0.32 & 6.34/0.32 \\
 & \texttt{prefix} & 5.85/0.53 & 5.79/0.74 & 7.67/0.94 & 6.32/0.63 & 8.22/1.11 & 5.04/1.43 & 6.08/0.63 & 6.50/0.58 \\
 & \texttt{random} & 10.79/0.80 & 9.62/0.96 & 5.01/0.42 & 4.80/0.26 & 7.32/0.32 & 4.77/0.42 & 5.92/0.42 & 5.86/0.37 \\

\midrule

\multicolumn{10}{c}{\textbf{News}} \\

\midrule

\multirow{2}{*}{\texttt{ablate}} & \texttt{lm} & 2.52/1.96 & 2.85/1.74 & 18.73/\textbf{14.42} & 18.31/12.08 & 1.56/0.86 & 0.98/0.44 & 0.85/\textbf{0.43} & 0.37/0.26 \\
& \texttt{in-situ} & 3.71/\textbf{2.21} & 3.43/1.85 & 2.37/1.12 & 2.84/1.58 & 2.37/0.84 & 0.76/0.33 & 1.38/0.34 & 0.58/0.21 \\
\cmidrule{1-10}
\multirow{3}{*}{\texttt{inject}} & \texttt{lm} & 2.64/0.32 & 2.75/0.26 & 18.63/2.68 & 16.28/1.97 & 1.53/0.33 & 1.91/0.27 & 1.20/0.16 & 1.68/0.05 \\
& \texttt{prefix} & 1.06/0.37 & 1.06/0.21 & 9.89/4.10 & 10.22/3.83 & 2.56/\textbf{0.87} & 1.36/0.44 & 1.36/0.21 & 2.25/0.37 \\
 & \texttt{random} & 3.27/0.90 & 3.01/0.69 & 4.70/1.58 & 5.30/1.37 & 1.96/0.55 & 1.20/0.22 & 0.73/0.26 & 1.41/0.42 \\

\midrule

\multicolumn{10}{c}{\textbf{GitHub}} \\

\midrule

\multirow{2}{*}{\texttt{ablate}} & \texttt{lm} & 8.70/6.62 & 8.34/6.08 & 8.31/5.89 & 6.73/4.93 & 11.08/\textbf{8.48} & 10.88/6.98 & 10.49/\textbf{7.45} & 7.89/5.63 \\
& \texttt{in-situ} & 2.78/1.85 & 2.55/1.56 & 4.62/2.35 & 3.13/1.20 & 4.03/2.42 & 1.80/1.11 & 3.79/2.37 & 1.74/1.22 \\
\cmidrule{1-10}
\multirow{3}{*}{\texttt{inject}} & \texttt{lm} & 13.01/7.08 & 13.41/\textbf{7.37} & 12.45/5.76 & 12.91/6.62 & 16.01/8.32 & 13.20/7.09 & 15.11/6.34 & 13.61/7.09 \\
& \texttt{prefix} & 3.69/2.13 & 3.40/2.65 & 15.36/9.21 & 14.77/\textbf{10.46} & 6.00/3.55 & 5.70/2.73 & 4.67/3.00 & 3.69/2.13 \\
& \texttt{random} & 2.53/1.44 & 1.96/1.44 & 7.22/4.30 & 8.74/5.56 & 2.50/1.05 & 2.38/1.05 & 2.54/1.38 & 2.36/1.33 \\

\bottomrule
\end{tabular}
}
\end{table*}

\subsection{Effectiveness and efficiency (RQ\textsubscript{2})}

\autoref{tab:combined_results} summarizes the effectiveness of the evaluated mutation configurations across three datasets and four model architectures. We report the ASR both \textbf{before} and \textbf{after} semantic verification using the entailment and perplexity thresholds ($\tau_{nli}$ and $\tau_{lps}$). This comparison provides insight into both the vulnerability of the evaluated models and the validity of the generated adversarial variants.
Across most evaluated settings, attribution-guided configurations combined with LLM-based semantic mutations generally increase post-verification ASR. 
\camera{In particular, LLM-driven ablation strategies expose substantially more verified failures than the in-situ method. }

A central observation emerging from \autoref{tab:combined_results} is the substantial discrepancy between pre-verification and post-verification ASR. In many cases, unconstrained perturbation strategies initially appear highly successful, but most generated variants fail semantic validation. For instance, under direct lexical ablation on SST-2 (DistilBERT), the pre-verification ASR reaches $17.57\%$, but decreases to only $1.71\%$ after semantic filtering. This large reduction indicates that many seemingly successful adversarial examples correspond to semantically invalid or label-altering perturbations rather than genuine robustness violations.

The semantic verification stage therefore plays a critical role in filtering invalid adversarial artifacts and ensuring that reported failures correspond to semantic-preserving inconsistencies. Under the default occlusion-guided configuration, LLM-based semantic mutations achieve a mean post-verification ASR of $3.70\%$ across evaluated datasets and architectures, with semantic ablation reaching $4.63\%$ and contextual injection $2.76\%$. In comparison, heuristic baselines achieve substantially lower verified ASR values, including In-situ Ablation ($1.48\%$), Prefix Injection ($2.20\%$), and Random Injection ($1.15\%$).
\camera{This combined reading shows that LLM-based ablation provides the most stable improvement across datasets, whereas LLM-based injection is more task-dependent, with stronger gains on GitHub issue classification than on SST-2 or AG News.}

\begin{table}[t]
\centering
\caption{
Pairwise comparisons using post-verification ASR. 
We report non-parametric 95\% bootstrap confidence intervals (CI), one-sided Wilcoxon test, and paired Cohen's $d_z$ as effect sizes.
%For each comparison, $n$ denotes the number of pairs.
%$\Delta$ denotes the paired mean difference (\emph{method} $-$ \emph{baseline}) in percentage points. 
}
\label{tab:rq2_pairwise_tests}
\small
\setlength{\tabcolsep}{4pt}
\resizebox{\textwidth}{!}{
\begin{tabular}{lllllrr}
\toprule
\textbf{Comparison} &
\textbf{$n$} &
\textbf{Method} &
\textbf{Baseline} &
\textbf{[95\% CI]} & %$\Delta$
\textbf{Wilcoxon $p$} & 
\textbf{$d_z$} \\
\midrule
LLM-Ablate vs. In-situ Ablation
& 24 & 4.27 & 1.48 &  \textbf{[1.51, 4.31]} & \textbf{<0.001} & 0.79\textsuperscript{M} \\

LLM-Inject vs. Random Injection
& 24 & 2.72 & 1.15 &  \textbf{[0.61, 2.66]} &\textbf{0.027}   & 0.60\textsuperscript{M} \\

LLM-Inject vs. Prefix Injection
& 24 & 2.72 & 2.20 & [-0.45, 1.56] & 0.634 & 0.20\textsuperscript{S} \\

Occlusion + LLM-Ablate vs. Random + LLM-Ablate
& 12 & 4.63 & 3.91 &\textbf{[0.30, 1.17]}& \textbf{0.005} & 0.89\textsuperscript{L} \\

Occlusion + LLM-Inject vs. Random + LLM-Inject
& 12 & 2.76 & 2.67 & [-0.22, 0.40] & 0.252 & 0.16\textsuperscript{N} \\
\bottomrule
\end{tabular}
}

\begin{tablenotes}[flushleft]
\footnotesize
\item The \textbf{bold} values indicate CI$_{95\%}(\Delta)$ is entirely above zero, or $p<0.05$. 
Effect sizes use paired Cohen's $d_z$:
\textsuperscript{N} negligible ($d<0.20$),
\textsuperscript{S} small ($0.20 \le d < 0.50$),
\textsuperscript{M} medium ($0.50 \le d < 0.80$),
and \textsuperscript{L} large ($d \ge 0.80$).
\end{tablenotes}

\end{table}

\changed{
Table~\ref{tab:rq2_pairwise_tests} reports the pairwise tests using post-verification ASR. The most consistent gain comes from LLM-Ablate, which significantly improves over in-situ deletion and also benefits from Occlusion-guided target selection. 
For injection, LLM-Inject significantly improves over Random Injection, but not significantly over the stronger Prefix Injection baseline. Thus, the evidence supports a robust advantage for
LLM-based ablation, while injection improvements are more setting-dependent.}

To assess computational practicality, we additionally measured the average execution time required to generate and verify a single metamorphic variant. Rule-based perturbation strategies remain highly efficient, requiring approximately $0.22$ seconds per generated attack on average. In contrast, LLM-based semantic mutations require approximately $4.71$ seconds due to the overhead introduced by local LLM inference and semantic verification. Although computationally more expensive, these configurations produce substantially higher proportions of semantically valid robustness violations.

\begin{tcolorbox}[boxrule=0pt,frame hidden,sharp corners,enhanced,borderline north={1pt}{0pt}{black},borderline south={1pt}{0pt}{black},boxsep=2pt,left=2pt,right=2pt,top=2.5pt,bottom=2pt]
\textbf{Effectiveness and efficiency (RQ\textsubscript{2})}: \textit{
LLM-based ablation provides the most robust gain over hard deletion, while LLM-based injection improves over random injection but not reliably over the stronger prefix baseline.
Semantic verification removes a large fraction of invalid adversarial artifacts, demonstrating that unconstrained perturbations substantially overestimate model brittleness.
}
\end{tcolorbox}

\subsection{Human validation (RQ\textsubscript{3})}

Table~\ref{tab:human_validation1} reports human validation results for the semantic gates. The NLI gate achieves a high pass precision of 94.7\%. However, a non-trivial portion of rejected variants are still judged valid by humans, suggesting that the gate is conservative and prioritizes precision over recall. This behavior is appropriate for our setting, where reducing false positives is more important than maximizing mutation
retention.
\begin{table}[t]
\centering
\caption{Confusion statistics between automatic gate decisions and human judgments. Pass precision is the human-positive rate among accepted variants.}
\label{tab:human_validation1}
\setlength{\tabcolsep}{4pt}
\resizebox{\textwidth}{!}{
\begin{tabular*}{\textwidth}{@{\extracolsep{\fill}}lcccccc}
\toprule
Comparison & TN & FP & FN & TP & Accuracy [\%] & Pass precision [\%]\\
\midrule
% Full gate vs. human valid & 139 & 38 & 79 & 127 & 69.5 & 61.7 \\
LPS gate vs. fluent & 29 & 13 & 66 & 209 & 75.1 & 76.0 \\
NLI gate vs. label-pres. & 42 & 66 & 10 & 179 & 74.4 & 94.7 \\
% Full gate vs. label-pres. & 42 & 77 & 10 & 168 & 70.7 & 94.4 \\
\bottomrule
\end{tabular*}
}
\end{table}

\begin{table}[h]
\centering
\caption{Human validation of semantic verification at configuration level. Human-valid is label-preserving and fluent.}
\label{tab:human_validation2}
\resizebox{\textwidth}{!}{
\begin{tabular}{lccccr}
\toprule
\textbf{Configuration}  & \textbf{full gate [\%]} & \textbf{human-valid [\%]} &
\textbf{label-pres. [\%]} & \textbf{fluent [\%]} & \textbf{gap [pp]} \\
\midrule
LLM-Ablate   & 80.6 & {74.1} & 89.8 & 95.5 & +6.5 \\
LLM-Inject     & 37.3 & {41.3} & 78.1 & 77.0 & -4.0 \\
Prefix-Inject & 47.8 & 23.9 & 81.1 & 42.1 & +23.9 \\
In-situ Ablate  & 37.5 & 19.3 & 73.7 & 43.5 & +18.2 \\
Random Inject  & 31.4 & 8.6  & 81.0 & 21.7 & +22.9 \\
\bottomrule
\end{tabular}
}
\end{table}

\autoref{tab:human_validation2} reports the human validation results compared to different mutation
configurations. LLM-Ablate achieves the highest human-valid rate and closely matches its automatic gate valid rate, suggesting that LLM-based ablation reliably preserves the intended metamorphic relation. For injection, LLM-Inject produces more human-valid and fluent variants than Prefix and Random Injection, even though its ASR advantage over Prefix is not statistically reliable in RQ2. This indicates that Prefix Injection is a strong failure-discovery baseline but often produces less natural variants. 
Overall, the semantic gate overestimates the quality of heuristic methods as the gaps are non-trivial, and LLM-based mutation produces more natural and usable test cases.

\begin{tcolorbox}[boxrule=0pt,frame hidden,sharp corners,enhanced,borderline north={1pt}{0pt}{black},borderline south={1pt}{0pt}{black},boxsep=2pt,left=2pt,right=2pt,top=2.5pt,bottom=2pt]
\textbf{Human validation (RQ\textsubscript{3})}: \textit{
The semantic gate acts as a high-precision but conservative validity filter. 
Human validation supports the use of LLM-based semantic rewriting, especially for producing
natural and label-preserving variants.
}
\end{tcolorbox}

\section{Threats to Validity}\label{sec:ttv}

\subsection{Internal validity}

\camera{The semantic verification thresholds may introduce outcome-aware calibration bias because they are selected from the observed mutation distributions. We mitigate this risk by using a single global threshold pair for all datasets, models, and configurations rather than tuning thresholds per method. We interpret the retained failures as verified candidate robustness failures rather than complete oracle-confirmed errors.
Furthermore, both automated semantic checks and human annotation provide evidence of label preservation rather than a complete oracle. Thus, some retained variants may still involve subtle semantic or label changes. We mitigate this risk through bidirectional entailment, fluency filtering, and human validation, and we interpret the results primarily at the aggregate level.}

LLM-based mutation generation is inherently stochastic and may be sensitive to prompting and decoding choices. To reduce this threat, we use fixed prompts, decoding parameters, and identical computational budgets across all configurations. Moreover, all baselines are implemented within the same framework and evaluated under identical settings, isolating the effect of the mutation strategy itself.

Finally, our approach relies on LLMs both for mutation generation and semantic verification, potentially introducing correlated biases. We mitigate this risk by separating generation and verification stages, combining multiple validation signals, and complementing automated filtering with manual inspection.

\subsection{External validity}

The generalizability of our findings may depend on the selected datasets, task semantics, mutation operators, and Judge LLM. To reduce overfitting conclusions to a specific setting, we evaluate multiple datasets and model families, including encoder-only, encoder--decoder, and decoder-only architectures.

Furthermore, our evaluation primarily relies on Attack Success Rate, which captures prediction inconsistencies but may not fully reflect the practical severity of failures. Finally, our mutation operators cover only a subset of possible semantic-preserving transformations; different perturbation spaces or future LLMs may lead to different behaviors.
% !TEX root =  paper.tex
\section{Qualitative Analysis}\label{sec:qualitative}

\begin{figure}[t]
  \centering
  \includegraphics[width=1\linewidth]{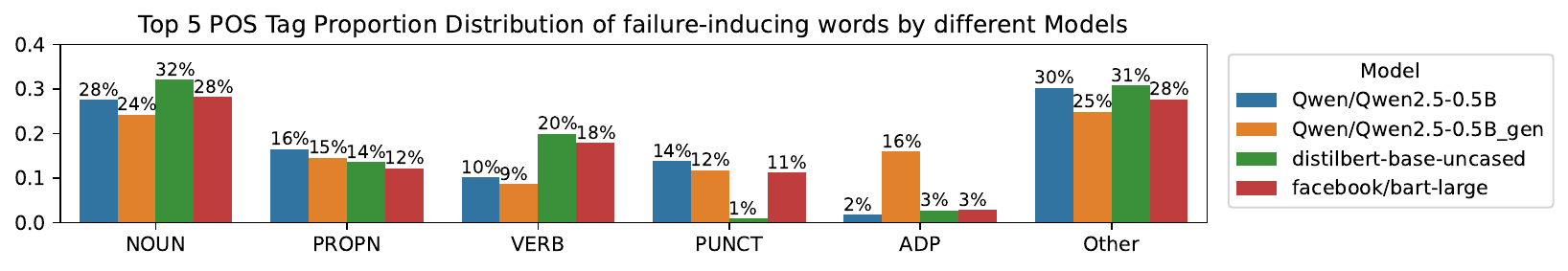}
  \caption{POS analysis of failure-inducing words.}
  \label{fig:pos}
\end{figure}

Beyond detecting failures, it is important to understand the patterns that trigger them. To this end, 
we conduct a qualitative and statistical analysis of 2320 verified failures produced by the best RQ\textsubscript{2} configuration across all three datasets, focusing on the failure-inducing words isolated by XAI and categorizing them by their Part-of-Speech (POS) tags.
As illustrated in Figure~\ref{fig:pos}, the distribution reveals systemic biases and distinct architectural vulnerabilities.

Across all four model architectures, Nouns (\texttt{NOUN}) consistently constitute the largest vulnerability surface.  
Furthermore, the nominal features, combining Nouns (\texttt{NOUN}) and Proper Nouns (\texttt{PROPN}),
contribute nearly half of the critical vulnerabilities. This strongly indicates that the SFT LMs frequently overfit to specific topical keywords or entities and bypass robust semantic reasoning.

To corroborate our statistical findings, we extract and inspect the specific failure-inducing tokens and their original contexts. This qualitative analysis unveils some distinct patterns of shortcut learning and semantic overfitting across the target models.
On the GitHub issue classification task, models like BART-Large exhibit a severe reliance on explicit, human-injected meta-words rather than comprehending the issue's technical description. For instance, ablating the literal word \texttt{"bug"} or \texttt{"question"} from the issue body consistently causes the model to misclassify the artifact, even when the surrounding text clearly describes an error trace (e.g., \textit{RuntimeError: source object number out of range'}) or a user query (e.g., \textit{May I ask if I can get a pull request...?'}). This reveals that the model acts largely as a shallow keyword-matcher, failing to generalize to issues where developers omit explicit category tags.
In news categorization, discriminative models (e.g., DistilBERT) frequently exploit Named Entities (\texttt{ENT\_TYPE=ORG}) as predictive shortcuts. We observe that models heavily rely on publisher acronyms (e.g., \texttt{"AFP"}, \texttt{"AP"}, \texttt{"Reuters"}) or dominant corporate entities (e.g., \texttt{"Apple"}, \texttt{"Pepsi"}) to determine the news category. For example, ablating \texttt{"(AFP)"} from a headline about European stocks causes the model's prediction to flip. This indicates a vulnerability where the SUT memorizes the data source distribution rather than reasoning about the semantic content of the headline.

\section{Discussion}\label{sec:discussion}

\head{Effectiveness of Attribution-Guided Mutation Selection}
The results of our study indicate that attribution-guided mutation selection substantially improves the discovery of verified robustness violations compared to unguided or purely heuristic perturbation strategies. Across datasets and architectures, configurations based on attribution signals generally achieve higher verified attack success rates and expose more semantically valid failures than random mutation baselines. These findings suggest that attribution methods provide an effective heuristic for identifying decision-relevant regions of the input space that are particularly sensitive to localized perturbations.

At the same time, the results also reveal differences across attribution techniques. Gradient-based and perturbation-based methods exhibit varying effectiveness depending on the model architecture and task, indicating that no single attribution strategy universally captures the most vulnerability-prone input regions. This variability further highlights the importance of systematically evaluating attribution methods within robustness testing settings rather than assuming uniform effectiveness across models.

\head{Impact of Semantic Verification}
A central finding of the study is that unconstrained adversarial generation substantially overestimates model brittleness. Many perturbations that successfully alter model predictions fail to preserve the original semantics or introduce severe linguistic degradation. Without semantic verification, these invalid mutations inflate attack success rates and provide misleading estimates of robustness weaknesses.

The human results also show why ASR alone is insufficient for comparing test-generation strategies.  
A configuration may be effective at inducing prediction flips while still producing variants that are unnatural, awkwardly inserted, or less useful for diagnosis. 
The results therefore demonstrate that semantic validation is essential for obtaining reliable robustness measurements in metamorphic testing settings.
Human validity further adds a complementary quality dimension, which is particularly important when comparing context-aware LLM mutations with stronger but more mechanical heuristics such as prefix-trigger insertion. The latter can be competitive in ASR, but human validation reveals that context-aware rewriting produces more natural and actionable variants.

\head{Evidence of Shortcut Reliance}
The discovered failures provide further evidence that task-specialized language models frequently rely on brittle lexical patterns and shortcut correlations. In many cases, small semantic-preserving edits affecting highly attributed tokens are sufficient to trigger prediction inconsistencies. This behavior is particularly pronounced for entity names, sentiment-bearing adjectives, and dataset-specific lexical indicators that strongly correlate with training labels.

These observations align with prior work on spurious correlations and annotation artifacts in NLP datasets, suggesting that strong in-distribution accuracy does not necessarily imply robust linguistic understanding. The results additionally indicate that attribution-guided perturbations can serve as a practical mechanism for systematically exposing such shortcut-driven behaviors.

\ifincludeRQThree
\head{Robustness Refinement}
The refinement experiments show that verified adversarial variants can be effectively reused as robustness-oriented training samples. Incorporating these variants during supervised fine-tuning consistently reduces vulnerability to metamorphic perturbations while preserving most of the original task accuracy. This suggests that semantically validated adversarial examples provide more useful refinement signals than unconstrained perturbations, which may otherwise introduce noisy or label-inconsistent training instances.

More generally, the results support the feasibility of iterative robustness evaluation settings in which detected failures are continuously incorporated into subsequent refinement stages to progressively improve model stability.
\fi

\head{Limitations}
The evaluated metamorphic relations focus primarily on lexical-level transformations. Although these perturbations already expose substantial robustness weaknesses, they do not capture higher-level linguistic phenomena such as discourse structure, reasoning consistency, or conversational context. Second, the semantic verification mechanisms rely on pretrained NLI and language models, which may themselves exhibit biases or verification errors. Third, the evaluation is limited to text classification tasks and moderate-sized task-specialized language models; additional studies are required to assess whether the observed findings generalize to larger generative models and more complex NLP tasks such as summarization, \camera{code generation, or question answering, which requires task-specific metamorphic relations and stronger validity oracles, which we leave for future work.}

Despite these limitations, the study demonstrates that combining attribution-guided mutation selection with semantic verification provides reliable and scalable evidence for robustness evaluation in specialized language models.
% !TEX root =  paper.tex
\section{Related Work}\label{sec:related}

\subsection{Test Generation for Deep Learning and Language Models}

Testing techniques for deep learning systems primarily focus on generating inputs that expose unstable, incorrect, or brittle model behaviors. Early work on adversarial testing demonstrated that small perturbations to the input can cause incorrect predictions while preserving human-perceived semantics. In NLP, character-level and word-level attacks such as HotFlip and DeepWordBug generate adversarial examples through gradient-based or heuristic perturbation strategies~\cite{ebrahimi-etal-2018-hotflip,gao2018blackboxgenerationadversarialtext}. 
Subsequent studies shifted toward semantically constrained attacks that aim to preserve fluency and meaning. For example, Alzantot et al.\ used genetic search over embedding neighbors to generate natural adversarial examples~\cite{alzantot2018generatingnaturallanguageadversarial}, while TextFooler prioritizes important tokens and replaces them with semantically similar alternatives~\cite{jin2020textfooler}. More recent approaches employ masked language models to generate context-aware substitutions and insertions that produce more fluent perturbations~\cite{li-etal-2020-bert-attack,garg-ramakrishnan-2020-bae}.

Beyond adversarial attacks, behavioral testing approaches assess model robustness through systematic input transformations designed to verify specific linguistic capabilities. CheckList introduced a taxonomy of behavioral tests based on invariance, directional expectation, and minimal functionality properties~\cite{ribeiro-etal-2020-beyond}. Other works construct contrast sets and adversarial benchmarks to reveal weaknesses that are not captured by standard evaluation datasets~\cite{gardner-etal-2020-evaluating,nie-etal-2020-adversarial,kiela-etal-2021-dynabench}. Collectively, these studies show that language models often fail under relatively simple linguistic variations despite strong benchmark performance.

However, many existing techniques rely on manually designed templates, heuristic perturbations, or unconstrained search procedures. Prior evaluations also report that adversarial generation methods frequently produce invalid examples that unintentionally alter semantics or degrade linguistic quality~\cite{morris2020textattack}. 
To address this problem, prior work has proposed semantic and naturalness checks for generated NLP test cases. AEON, for example, evaluates whether generated NLP test cases
preserve semantic similarity and natural language quality~\cite{huang2022aeon}.
Metamorphic Testing (MT) has also been adapted to NLP through predefined metamorphic relations that verify prediction consistency under semantic-preserving transformations~\cite{chen1998metamorphic,segura2016survey,11185922}. Nevertheless, many existing MT techniques still rely on random or weakly guided perturbations and provide limited guarantees regarding semantic validity.

This study is motivated by the observation that these components are often evaluated in isolation. \camera{Our novelty lies in empirically investigating the combination of attribution-guided mutation selection, LLM-based semantic transformations, and automated semantic verification for robustness testing of task-specialized language models.}

\subsection{Detecting Spurious Correlations in Language Models}

A growing body of research investigates how language models exploit superficial patterns and spurious correlations present in training data. Adversarial evaluation studies have shown that models frequently rely on shallow lexical cues instead of robust linguistic reasoning. Universal adversarial triggers, for instance, demonstrate that short token sequences can systematically manipulate model predictions across many inputs~\cite{wallace2019triggers}. Similarly, adversarial NLI benchmarks reveal that models often exploit annotation artifacts and dataset-specific heuristics rather than genuine semantic understanding~\cite{nie-etal-2020-adversarial}.

Explainable AI techniques have also been widely used to diagnose such behaviors by identifying which input regions contribute most strongly to model predictions~\cite{zhao2024explainability}. Attribution methods such as Integrated Gradients~\cite{sundararajan2017integrated}, attention-based analysis~\cite{abnar2020attention}, and relevance propagation approaches~\cite{chefer2021transformer} provide insight into the evidence used by neural models during inference~\cite{2026-Chen-arXiv}. These attribution signals have subsequently been incorporated into adversarial example generation and robustness analysis pipelines~\cite{jin2020textfooler,bayer-etal-2024-xai}.

Despite these advances, explainability methods are predominantly used as post-hoc diagnostic tools rather than as systematic mechanisms for generating semantically controlled robustness tests. 
This study builds on prior findings by empirically evaluating whether attribution signals can effectively guide metamorphic mutation selection and improve the generation of valid failure-inducing test cases. 
We also analyze the resulting failures to characterize recurring shortcut patterns, such as over-reliance on named entities and topical keywords. In doing so, the study connects attribution-based diagnosis with metamorphic robustness testing.
The results provide further evidence on the relationship between attribution-guided perturbations, semantic validity, and brittle decision patterns in specialized language models. 

% !TEX root =  paper.tex
\section{Conclusions}\label{sec:conclusions}

In this paper, we presented an empirical study on XAI-guided metamorphic robustness testing for task-specialized language models. The study investigated how attribution-guided mutation prioritization, LLM-based semantic perturbations, and semantic verification mechanisms influence the generation of valid failure-inducing test variants without requiring test oracles.

Our evaluation across multiple datasets, architectures, and classification paradigms shows that combining explanation-guided selection with LLM-based mutation consistently exposes more robustness violations than unguided and rule-based baselines. 
Semantic verification filters a substantial portion of invalid adversarial artifacts. The human study further shows that the semantic gate is high-precision but conservative, making it suitable for robustness testing settings where false positives are costly. The verified failures also reveal shortcut behaviors, including sensitivity to named entities, lexical indicators, and formatting cues.

%The study also shows that semantic verification mechanisms based on bidirectional entailment and perplexity filtering substantially improve mutation validity, enabling more reliable robustness assessment than purely heuristic perturbation strategies. In addition, the generated verified variants can support robustness-oriented refinement with limited degradation on the original task distribution.
% Our results show that explanation- and LLM-guided mutation selection improve verified failure discovery, while semantic verification and human validation substantially reduce the risk of reporting invalid adversarial artifacts. The human study indicates that the semantic gate is high-precision but conservative, making it suitable for robustness testing settings where false positives are costly.

Future work will investigate richer metamorphic transformations beyond lexical edits, including structural, discourse-level, and conversational perturbations, as well as adaptive robustness testing settings in which discovered failures iteratively guide subsequent model refinement and evaluation.

\section{Data Availability Statement}

We provide an anonymous replication package with our experimental pipeline and processed datasets~\cite{replication-package}. A public DOI is not released at submission time to avoid premature dissemination; the repository will be archived with a DOI upon acceptance.

% \Andrea{references are too many and too long, the paper's limits are 20 pages (17 + 3)}

%\balance
\bibliography{bibi} 

@article{DistilBERTAD,
	title        = {DistilBERT, a distilled version of BERT: smaller, faster, cheaper and lighter},
	author       = {Victor Sanh and Lysandre Debut and Julien Chaumond and Thomas Wolf},
	year         = 2019,
	journal      = {ArXiv},
	volume       = {abs/1910.01108}
}

@misc{qwen,
	title        = {Qwen2.5: A Party of Foundation Models},
	author       = {Qwen Team},
	year         = 2024,
	month        = {September},
	url          = {https://qwenlm.github.io/blog/qwen2.5/}
}

@misc{2026-Chen-arXiv,
	title        = {Feature-Aware Test Generation for Deep Learning Models},
	author       = {Chen, Xingcheng and Weissl, Oliver and Stocco, Andrea},
	year         = 2026,
	url          = {https://arxiv.org/abs/2601.14081},
	eprint       = {},
	archiveprefix = {arXiv},
	primaryclass = {cs.SE}
}

@inproceedings{bart,
	title        = {BART: Denoising Sequence-to-Sequence Pre-training for Natural Language Generation, Translation, and Comprehension},
	author       = {Lewis, Mike  and Liu, Yinhan  and Goyal, Naman  and Ghazvininejad, Marjan  and Mohamed, Abdelrahman  and Levy, Omer  and Stoyanov, Veselin  and Zettlemoyer, Luke},
	year         = 2020,
	month        = jul,
	booktitle    = {Proc. ACL 2020},
	publisher    = {Association for Computational Linguistics},
	pages        = {7871--7880},
	doi          = {10.18653/v1/2020.acl-main.703},
}

@inproceedings{vaswani2017attention,
	title        = {Attention is All you Need},
	author       = {Vaswani, Ashish and Shazeer, Noam and Parmar, Niki and Uszkoreit, Jakob and Jones, Llion and Gomez, Aidan N and Kaiser, Lukasz and Polosukhin, Illia},
	year         = 2017,
	journal      = {Advances in Neural Information Processing Systems},
	booktitle    = {Advances in Neural Information Processing Systems},
	publisher    = {Curran Associates, Inc.},
	volume       = 30,
	pages        = {},
}

@inproceedings{devlin2019bert,
	title        = {BERT: Pre-training of Deep Bidirectional Transformers for Language Understanding},
	author       = {Devlin, Jacob  and Chang, Ming-Wei  and Lee, Kenton  and Toutanova, Kristina},
	year         = 2019,
	month        = jun,
	booktitle    = {Proceedings of the 2019 Conference of the North {A}merican Chapter of the Association for Computational Linguistics: Human Language Technologies, Volume 1 (Long and Short Papers)},
	publisher    = {Association for Computational Linguistics},
	address      = {Minneapolis, Minnesota},
	pages        = {4171--4186},
	doi          = {10.18653/v1/N19-1423}
}

@article{raffel2020t5,
	title        = {Exploring the Limits of Transfer Learning with a Unified Text-to-Text Transformer},
	author       = {Colin Raffel and Noam Shazeer and Adam Roberts and Katherine Lee and Sharan Narang and Michael Matena and Yanqi Zhou and Wei Li and Peter J. Liu},
	year         = 2020,
	journal      = {Journal of Machine Learning Research},
	volume       = 21,
	number       = 140,
	pages        = {1--67}
}

@inproceedings{sundararajan2017integrated,
	title        = {Axiomatic attribution for deep networks},
	author       = {Sundararajan, Mukund and Taly, Ankur and Yan, Qiqi},
	year         = 2017,
	booktitle    = {Proceedings of the 34th International Conference on Machine Learning - Volume 70},
	location     = {Sydney, NSW, Australia},
	publisher    = {JMLR.org},
	series       = {ICML'17},
	pages        = {3319–3328},
	numpages     = 10
}

@inproceedings{ribeiro2016lime,
	title        = {"Why Should I Trust You?": Explaining the Predictions of Any Classifier},
	author       = {Marco Tulio Ribeiro and Sameer Singh and Carlos Guestrin},
	year         = 2016,
	month        = jun,
	booktitle    = {Proceedings of the 2016 Conference of the North {A}merican Chapter of the Association for Computational Linguistics: Demonstrations},
	publisher    = {Association for Computational Linguistics},
	address      = {San Diego, California},
	pages        = {97--101},
}

@inproceedings{zeiler2014occlusion,
	title        = {Visualizing and Understanding Convolutional Networks},
	author       = {Matthew D. Zeiler and Rob Fergus},
	year         = 2014,
	booktitle    = {Computer Vision -- ECCV 2014},
	publisher    = {Springer International Publishing},
	address      = {Cham},
	pages        = {818--833},
	isbn         = {978-3-319-10590-1}
}

@article{chen1998metamorphic,
	title        = {Metamorphic Testing: A New Approach for Generating Next Test Cases},
	author       = {Tsong Yueh Chen and Siu Cheung and Siu-Ming Yiu},
	year         = 1998,
	journal      = {Technical Report HKUST-CS98-01}
}

@article{segura2016survey,
	title        = {A Survey on Metamorphic Testing},
	author       = {Sergio Segura and Gordon Fraser and Ana B. Sanchez and Antonio Ruiz-Cortes},
	year         = 2016,
	journal      = {IEEE Transactions on Software Engineering},
	volume       = 42,
	number       = 9,
	pages        = {805--824},
}

@inproceedings{abnar2020attention,
	title        = {Quantifying Attention Flow in Transformers},
	author       = {Abnar, Samira and Zuidema, Willem},
	year         = 2020,
	booktitle    = {Proc. ACL 2020},
	pages        = {4190--4197},
	doi          = {10.18653/v1/2020.acl-main.385}
}

@inproceedings{chefer2021transformer,
	title        = {Transformer Interpretability Beyond Attention Visualization},
	author       = {Chefer, Hila and Gur, Shir and Wolf, Lior},
	year         = 2021,
	booktitle    = {2021 IEEE/CVF Conference on Computer Vision and Pattern Recognition (CVPR)},
	pages        = {782--791},
	doi          = {10.1109/CVPR46437.2021.00084}
}

@inproceedings{williams2018mnli,
	title        = {A Broad-Coverage Challenge Corpus for Sentence Understanding through Inference},
	author       = {Williams, Adina and Nangia, Nikita and Bowman, Samuel},
	year         = 2018,
	booktitle    = {Proc. NAACL-HLT 2018},
}

@article{liu2019roberta,
	title        = {RoBERTa: A Robustly Optimized {BERT} Pretraining Approach},
	author       = {Liu, Yinhan and Ott, Myle and Goyal, Naman and Du, Jingfei and Joshi, Mandar and Chen, Danqi and Levy, Omer and Lewis, Mike and Zettlemoyer, Luke and Stoyanov, Veselin},
	year         = 2019,
	journal      = {CoRR},
	volume       = {abs/1907.11692}
}

@article{jozefowicz2016exploring,
	title        = {Exploring the Limits of Language Modeling},
	author       = {Jozefowicz, Rafal and Vinyals, Oriol and Schuster, Mike and Shazeer, Noam and Wu, Yonghui},
	year         = 2016,
	journal      = {arXiv preprint arXiv:1602.02410}
}

@inproceedings{nlbse2022,
	title        = {NLBSE'22 Tool Competition},
	author       = {Kallis, Rafael and Chaparro, Oscar and Di Sorbo, Andrea and Panichella, Sebastiano},
	year         = 2022,
	booktitle    = {2022 IEEE/ACM 1st International Workshop on Natural Language-Based Software Engineering (NLBSE)},
	publisher    = {IEEE},
	pages        = {25--28},
	doi          = {10.1145/3528588.3528664}
}

@inproceedings{zhang2015agnews,
	title        = {Character-level Convolutional Networks for Text Classification},
	author       = {Zhang, Xiang and Zhao, Junbo and LeCun, Yann},
	year         = 2015,
	booktitle    = {Proceedings of the 29th International Conference on Neural Information Processing Systems - Volume 1},
	pages        = {649--657}
}

@inproceedings{socher2013sst,
	title        = {Recursive Deep Models for Semantic Compositionality Over a Sentiment Treebank},
	author       = {Socher, Richard and Perelygin, Alex and Wu, Jean and Chuang, Jason and Manning, Christopher D. and Ng, Andrew and Potts, Christopher},
	year         = 2013,
	booktitle    = {EMNLP},
	publisher    = {Association for Computational Linguistics},
}

@inproceedings{wallace2019triggers,
	title        = {Universal Adversarial Triggers for Attacking and Analyzing NLP},
	author       = {Wallace, Eric and Feng, Shi and Kandpal, Nikhil and Gardner, Matt and Singh, Sameer},
	year         = 2019,
	booktitle    = {Proceedings of EMNLP-IJCNLP}
}

@inproceedings{li2016erasure,
	title        = {Understanding Neural Networks through Representation Erasure},
	author       = {Li, Jiwei and Monroe, Will and Jurafsky, Dan},
	year         = 2016,
	booktitle    = {Proceedings of arXiv:1612.08220}
}

@inproceedings{li2019textbugger,
	title        = {TextBugger: Generating Adversarial Text Against Real-world Applications},
	author       = {Li, Jinfeng and Ji, Shouling and Du, Tianyu and Li, Bo and Wang, Ting},
	year         = 2019,
	booktitle    = {Proceedings of NDSS}
}

@inproceedings{kneed,
	title        = {Finding a "Kneedle" in a Haystack: Detecting Knee Points in System Behavior},
	author       = {Satopaa, Ville and Albrecht, Jeannie and Irwin, David and Raghavan, Barath},
	year         = 2011,
	booktitle    = {2011 31st International Conference on Distributed Computing Systems Workshops},
	volume       = {},
	number       = {},
	pages        = {166--171}
}

@inproceedings{gardner-etal-2020-evaluating,
	title        = {Evaluating Models' Local Decision Boundaries via Contrast Sets},
	author       = {Gardner, Matt  and others},
	year         = 2020,
	month        = nov,
	booktitle    = {Findings of the Association for Computational Linguistics: EMNLP 2020},
	publisher    = {Association for Computational Linguistics},
	pages        = {1307--1323},
}

@inproceedings{ribeiro-etal-2020-beyond,
	title        = {Beyond Accuracy: Behavioral Testing of {NLP} Models with {C}heck{L}ist},
	author       = {Ribeiro, Marco Tulio  and Wu, Tongshuang  and Guestrin, Carlos  and Singh, Sameer},
	year         = 2020,
	booktitle    = {Proc. ACL 2020},
	publisher    = {Association for Computational Linguistics}
}

@inproceedings{nie-etal-2020-adversarial,
	title        = {Adversarial {NLI}: A New Benchmark for Natural Language Understanding},
	author       = {Nie, Yixin  and Williams, Adina  and Dinan, Emily  and Bansal, Mohit  and Weston, Jason  and Kiela, Douwe},
	year         = 2020,
	month        = jul,
	booktitle    = {Proc. ACL 2020},
	publisher    = {Association for Computational Linguistics},
	pages        = {4885--4901},
}

@inproceedings{kiela-etal-2021-dynabench,
	title        = {Dynabench: Rethinking Benchmarking in {NLP}},
	author       = {Kiela, Douwe  and others},
	year         = 2021,
	month        = jun,
	booktitle    = {Proceedings of the 2021 Conference of the North American Chapter of the Association for Computational Linguistics: Human Language Technologies},
	pages        = {4110--4124},
}

@article{
liang2023holisticevaluationlanguagemodels,
title={Holistic Evaluation of Language Models},
author={Percy Liang and others},
journal={Transactions on Machine Learning Research},
issn={2835-8856},
year={2023},
note={Featured Certification, Expert Certification}
}

@inproceedings{ebrahimi-etal-2018-hotflip,
	title        = {{H}ot{F}lip: White-Box Adversarial Examples for Text Classification},
	author       = {Ebrahimi, Javid  and Rao, Anyi  and Lowd, Daniel  and Dou, Dejing},
	year         = 2018,
	month        = jul,
	booktitle    = {Proceedings of the 56th Annual Meeting of the Association for Computational Linguistics (Volume 2: Short Papers)},
	publisher    = {Association for Computational Linguistics},
	address      = {Melbourne, Australia},
	pages        = {31--36},
	doi          = {10.18653/v1/P18-2006},
}

@INPROCEEDINGS{gao2018blackboxgenerationadversarialtext,
  author={Gao, Ji and Lanchantin, Jack and Soffa, Mary Lou and Qi, Yanjun},
  booktitle={2018 IEEE Security and Privacy Workshops (SPW)}, 
  title={Black-Box Generation of Adversarial Text Sequences to Evade Deep Learning Classifiers}, 
  year={2018},
  volume={},
  number={},
  pages={50-56},
  doi={10.1109/SPW.2018.00016}}

@inproceedings{alzantot2018generatingnaturallanguageadversarial,
    title = "Generating Natural Language Adversarial Examples",
    author = "Alzantot, Moustafa  and
      Sharma, Yash  and
      Elgohary, Ahmed  and
      Ho, Bo-Jhang  and
      Srivastava, Mani  and
      Chang, Kai-Wei",
    booktitle = "Proceedings of the 2018 Conference on Empirical Methods in Natural Language Processing",
    month = oct # "-" # nov,
    year = "2018",
    address = "Brussels, Belgium",
    publisher = "Association for Computational Linguistics",
    pages = "2890--2896",
}

@inproceedings{li-etal-2020-bert-attack,
	title        = {{BERT}-{ATTACK}: Adversarial Attack Against {BERT} Using {BERT}},
	author       = {Li, Linyang  and Ma, Ruotian  and Guo, Qipeng  and Xue, Xiangyang  and Qiu, Xipeng},
	year         = 2020,
	month        = nov,
	booktitle    = {Proceedings of the 2020 Conference on Empirical Methods in Natural Language Processing (EMNLP)},
	publisher    = {Association for Computational Linguistics},
	pages        = {6193--6202},
}

@inproceedings{garg-ramakrishnan-2020-bae,
	title        = {{BAE}: {BERT}-based Adversarial Examples for Text Classification},
	author       = {Garg, Siddhant  and Ramakrishnan, Goutham},
	year         = 2020,
	month        = nov,
	booktitle    = {EMNLP},
	publisher    = {Association for Computational Linguistics},
	pages        = {6174--6181},
}

@inproceedings{bayer-etal-2024-xai,
	title        = {{XAI}-Attack: Utilizing Explainable {AI} to Find Incorrectly Learned Patterns for Black-Box Adversarial Example Creation},
	author       = {Bayer, Markus  and Neiczer, Markus  and Samsinger, Maximilian  and Buchhold, Bj{\"o}rn  and Reuter, Christian},
	year         = 2024,
	booktitle    = {LREC-COLING},
	publisher    = {ELRA and ICCL},
	address      = {Torino, Italia},
}

@inproceedings{wang2019gluemultitaskbenchmarkanalysis,
    title = "{GLUE}: A Multi-Task Benchmark and Analysis Platform for Natural Language Understanding",
    author = "Wang, Alex  and others",
    booktitle = "Proceedings of the 2018 {EMNLP} Workshop {B}lackbox{NLP}: Analyzing and Interpreting Neural Networks for {NLP}",
    month = nov,
    year = "2018",
    address = "Brussels, Belgium",
    publisher = "Association for Computational Linguistics",
    doi = "10.18653/v1/W18-5446",
    pages = "353--355",
}

@article{zhao2024explainability,
author = {Zhao, Haiyan and Chen, Hanjie and Yang, Fan and Liu, Ninghao and Deng, Huiqi and Cai, Hengyi and Wang, Shuaiqiang and Yin, Dawei and Du, Mengnan},
title = {Explainability for Large Language Models: A Survey},
year = {2024},
issue_date = {April 2024},
publisher = {Association for Computing Machinery},
address = {New York, NY, USA},
volume = {15},
number = {2},
issn = {2157-6904},
url = {https://doi.org/10.1145/3639372},
journal = {ACM Trans. Intell. Syst. Technol.},
month = feb,
articleno = {20},
numpages = {38}
}

@INPROCEEDINGS{11185922,
  author={Cho, Steven and Ruberto, Stefano and Terragni, Valerio},
  booktitle={2025 IEEE International Conference on Software Maintenance and Evolution (ICSME)}, 
  title={Metamorphic Testing of Large Language Models for Natural Language Processing}, 
  year={2025},
  volume={},
  number={},
  pages={174-186},}

@misc{replication-package,
	title        = {Replication Package},
	year         = 2026,
	doi          = {},
	url          = {https://github.com/ast-fortiss-tum/lexcheck},
	key          = {\tool},
	howpublished = {}
}

@misc{he2021debertav3,
      title={DeBERTaV3: Improving DeBERTa using ELECTRA-Style Pre-Training with Gradient-Disentangled Embedding Sharing}, 
      author={Pengcheng He and Jianfeng Gao and Weizhu Chen},
      year={2021},
      eprint={2111.09543},
      archivePrefix={arXiv},
      primaryClass={cs.CL}
}

@inproceedings{jin2020textfooler,
  title={Is BERT Really Robust? A Strong Baseline for Natural Language Attack on Text Classification and Entailment},
  author={Jin, Di and Jin, Zhijing and Zhou, Joey Tianyi and Szolovits, Peter},
  booktitle={AAAI Conference on Artificial Intelligence},
  year={2019}
}

@inproceedings{morris2020textattack,
  title={TextAttack: A Framework for Adversarial Attacks, Data Augmentation, and Adversarial Training in NLP},
  author={Morris, John X and Lifland, Eli and Yoo, Jin Yong and Grigsby, Jake and Jin, Di and Qi, Yanjun},
  booktitle={Conference on Empirical Methods in Natural Language Processing},
  year={2020}
}

@inproceedings{qi2021openattack,
  title={{Openattack: An open-source textual adversarial attack toolkit}},
  author={Zeng, Guoyang and Qi, Fanchao and Zhou, Qianrui and Zhang, Tingji and Hou, Bairu and Zang, Yuan and Liu, Zhiyuan and Sun, Maosong},
  booktitle={Proc. ACL-IJCNLP 2021: System Demonstrations},
  pages={363--371},
  year={2021},
}

@article{radford2019language,
  title={Language models are unsupervised multitask learners},
  author={Alec Radford and Jeff Wu and Rewon Child and David Luan and Dario Amodei and Ilya Sutskever},
  journal={OpenAI blog},
  volume={1},
  number={8},
  pages={9},
  year={2019}
}

@article{10.1145/3641289,
author = {Chang, Yupeng and others},
title = {A Survey on Evaluation of Large Language Models},
year = {2024},
issue_date = {June 2024},
publisher = {Association for Computing Machinery},
address = {New York, NY, USA},
volume = {15},
number = {3},
issn = {2157-6904},
url = {https://doi.org/10.1145/3641289},
doi = {10.1145/3641289},
journal = {ACM Trans. Intell. Syst. Technol.},
month = mar,
articleno = {39},
numpages = {45}
}

@inproceedings{mudrakarta2018did,
    title = "Did the Model Understand the Question?",
    author = "Mudrakarta, Pramod Kaushik  and
      Taly, Ankur  and
      Sundararajan, Mukund  and
      Dhamdhere, Kedar",
    booktitle = "Proceedings of the 56th Annual Meeting of the Association for Computational Linguistics (Volume 1: Long Papers)",
    month = jul,
    year = "2018",
    address = "Melbourne, Australia",
    publisher = "Association for Computational Linguistics",
    pages = "1896--1906",
}

@inproceedings{mosbach2020on,
  title={On the Stability of Fine-tuning {BERT}: Misconceptions, Explanations, and Strong Baselines},
  author={Mosbach, Marius and Andriushchenko, Maksym and Klakow, Dietrich},
  booktitle={International Conference on Learning Representations (ICLR)},
  year={2021}
}

@inproceedings{sorokin2008utility,
	title        = {Utility data annotation with amazon mechanical turk},
	author       = {Sorokin, Alexander and Forsyth, David},
	year         = 2008,
	booktitle    = {2008 IEEE computer society conference on computer vision and pattern recognition workshops},
	pages        = {1--8},
	organization = {IEEE}
}

@inproceedings{huang2022aeon,
author = {Huang, Jen-tse and Zhang, Jianping and Wang, Wenxuan and He, Pinjia and Su, Yuxin and Lyu, Michael R.},
title = {AEON: a method for automatic evaluation of NLP test cases},
year = {2022},
isbn = {9781450393799},
publisher = {Association for Computing Machinery},
address = {New York, NY, USA},
pages = {202–214},
numpages = {13},
location = {Virtual, South Korea},
series = {ISSTA 2022}
}

@book{cohen1988statistical,
	title        = {Statistical power analysis for the behavioral sciences},
	author       = {Cohen, Jacob},
	year         = 1988,
	publisher    = {L. Erlbaum Associates},
	address      = {Hillsdale, N.J},
	isbn         = {978-1-134-74270-7}
}

@article{Wilcoxon1945,
	title        = {Individual Comparisons by Ranking Methods},
	author       = {Frank Wilcoxon},
	year         = 1945,
	month        = dec,
	journal      = {Biometrics Bulletin},
	publisher    = {{JSTOR}},
	volume       = 1,
	number       = 6,
	pages        = 80,
	doi          = {10.2307/3001968},
	url          = {https://doi.org/10.2307/3001968}
}

@article{tibshirani1993bootstrap,
  title={An introduction to the bootstrap},
  author={Tibshirani, Robert J and Efron, Bradley},
  journal={Monographs on statistics and applied probability},
  volume={57},
  number={1},
  pages={1--436},
  year={1993}
}

@INPROCEEDINGS{METAL,
  author={Hyun, Sangwon and Guo, Mingyu and Babar, M. Ali},
  booktitle={ICST}, 
  title={METAL: Metamorphic Testing Framework for Analyzing Large-Language Model Qualities}, 
  year={2024},
  volume={},
  number={},
  pages={117-128},
  doi={10.1109/ICST60714.2024.00019}}
\end{document}